\documentclass[11pt,a4paper]{article}

\usepackage{amssymb}
\usepackage{citesort}
\usepackage{amsmath}
\usepackage{url}

\def\de{\delta}

\def\na{\nabla}

\def\pa{\partial}
\def\fr{\frac}

\def\al{\alpha}

\begin{document}
\vspace*{1.0cm}
\noindent
{\bf
{\large
\begin{center}
A new pilot-wave model for quantum field theory
\end{center}
}
}

\vspace*{.5cm}
\begin{center}
W.\ Struyve, H.\ Westman
\end{center}

\begin{center}
Perimeter Institute for Theoretical Physics \\
31 Caroline Street North, Waterloo, Ontario N2L 2Y5, Canada \\
E--mail: wstruyve@perimeterinstitute.ca, hwestman@perimeterinstitute.ca
\end{center}

\begin{abstract}
\noindent
We present a way to construct a pilot-wave model for quantum field theory. The idea is to introduce beables corresponding only to the bosonic degrees of freedom and not to the fermionic degrees of freedom of the quantum state. We illustrate this idea for quantum electrodynamics. The beables will be field beables corresponding to the electromagnetic field and they will be introduced in a similar way to that of Bohm's model for the free electromagnetic field. Our approach is analogous to the situation in non-relativistic quantum theory, where Bell treated spin not as a beable but only as a property of the wavefunction.
\end{abstract}

\bibliographystyle{unsrt}

\section{Introduction}
Already in his seminal paper in 1952, Bohm presented a pilot-wave interpretation for the free electromagnetic field \cite{bohm2}. The beables in his pilot-wave model were fields. Similar models can be constructed for the other bosonic fields that are present in the `standard model' for high energy physics, i.e.\ the (electro-)weak interaction field, strong interaction field and Higgs field \cite{valentini04,valentini0502,struyve052}. 

On the other hand, for fermionic quantum field theory, no good pilot-wave model in terms of field beables has been presented yet. There are two attempts to construct a pilot-wave model for fermionic field theory with fields as beables, one by Holland \cite{holland,holland881} and another one by Valentini \cite{valentini92,valentini96}, but both of them have problems \cite{struyve051,struyve052}. So far, particle beables seem more successful for fermionic quantum field theory. Bell presented a model for quantum field theory on a lattice \cite{bell86}, where the beables are the fermion numbers at each lattice point. Bell's model differs from the usual pilot-wave program in the fact that it is indeterministic. However, Bell expected that the indeterminism would disappear in the continuum limit. Work by Colin \cite{colin031,colin032,colin033} seems to confirm Bell's expectation. On the other hand D\"urr {\em et al.} have developed a continuum version of Bell's model which is stochastic \cite{durr02,durr031,durr032,tumulka03,durr04}.

In this paper we present an alternative approach to a pilot-wave model for quantum field theory. Making use of the fact that all fermionic fields are gauge coupled to bosonic fields, we will argue that it is sufficient to introduce beables corresponding only to the bosonic degrees of freedom. No beables need to be introduced corresponding to the fermionic degrees of freedom. We illustrate this idea for quantum electrodynamics. The beables will be field beables corresponding to the electromagnetic field and they will be introduced in a similar way to that of Bohm's model for the free electromagnetic field. In this way we obtain a deterministic pilot-wave model. The strategy of not associating beables with all the degrees of freedom of the quantum state has been exploited before in some pilot-wave models. There is for example Bell's model for non-relativistic spin-1/2 particles, where no beables are associated with the spin degrees of freedom. 

In the next section we start by recalling the pilot-wave theory for non-relativistic quantum systems that was presented by de Broglie and Bohm. We will thereby emphasize the importance of effective collapse in pilot-wave theory in order to show the empirical equivalence with standard quantum theory. When we present our pilot-wave model for quantum field theory, we will use a similar notion of effective collapse in order to show that our model reproduces the empirical predictions of standard quantum theory. Before we present our model in Section \ref{themodel}, we describe models in Section \ref{similarmodels}, which are similar to our model in the sense that they do not associate beables to every degree of freedom in the wavefunction.

\section{Pilot-wave theory and the empirical equivalence with quantum theory}
In the pilot-wave theory for non-relativistic quantum systems by de Broglie and Bohm \cite{debroglie28,bohm1,bohm2}, the complete description of a quantum system is provided by its wavefunction and by point-particles which have definite positions at all times. In order to make a clear distinction between the notion of particles in quantum theory and the particles that are introduced as additional variables in pilot-wave theory, we will refer to the latter as {\it particle beables} \cite{bell87}. For a system with a wavefunction $\psi({\bf x}_1,\dots,{\bf x}_N,t)$, with $\psi=|\psi|\exp(iS/\hbar)$, the possible trajectories $({\bf x}_1(t),\dots,{\bf x}_N(t))$ of the particles beables are solutions to the guiding equations
\begin{equation}
\fr{d {\bf x}_k}{dt}= \frac{1}{m_k} {\boldsymbol {\nabla}}_k S\,,
\label{1}
\end{equation}
with $k=1,\dots,N$. If one considers a quantum measurement on an ensemble of identically prepared systems (all described by the same wavefunction), then, as is well known, pilot-wave theory reproduces the statistics of quantum theory if the particles beables are distributed according to $|\psi({\bf x}_1,\dots,{\bf x}_N,t)|^2$ over the ensemble. This particular distribution is is called the {\it equilibrium distribution} \cite{valentini92,valentini,durr92}. Throughout this paper we assume equilibrium distributions. Quantum equilibrium can be justified by applying statistical arguments to pilot-wave theory \cite{valentini92,valentini,valentini04,valentini042,durr92}.

However, in order to have a good pilot-wave model, i.e.\ a model which is empirically equivalent to quantum theory, the requirement of quantum equilibrium is necessary but not sufficient. The pilot-wave model also needs to exhibit {\em effective collapse}, at least in situations where you expect {\em ordinary} collapse to occur in standard quantum theory,{\footnote{By `standard quantum theory' we mean the Dirac-von Neumann formulation in which the wavefunction evolves according to the Schr\"odinger equation until collapse occurs. It is sufficient to compare the empirical predictions of pilot-wave theory to the Dirac-von Neumann formulation of standard quantum theory, because `standard' formulations of quantum theory, like Bohr's, agree with the Dirac-von Neumann formulation at the empirical level.}} for example in quantum measurement situations or when we have a superposition of macroscopically distinct states. Hence, although the collapse rule of standard quantum theory is not a part of the pilot-wave formulation, it should arise as an emergent phenomenon and this phenomenon is called effective collapse. 

Let us first of all explain what effective collapse is in the pilot-wave theory for non-relativistic quantum systems. Suppose we have a system which is described by a superposition 
\begin{equation}
\psi = \psi_1 + \psi_2\,.
\label{1.1}
\end{equation}
The wavefunctions $\psi_1$ and $\psi_2$ are said to be non-overlapping at a certain time $t_0$ if 
\begin{equation}
\psi_1 ({\bf x}_1,\dots,{\bf x}_N,t_0)\psi_2({\bf x}_1,\dots,{\bf x}_N,t_0) = 0\,,  \quad \forall ({\bf x}_1,\dots,{\bf x}_N)\in {\mathbb R}^{3N}\,.
\label{1.101}
\end{equation}
If $\psi_1$ and $\psi_2$ are non-overlapping for the time interval $I=[t_0,+\infty)$, then, for $t \in I$, the density $|\psi|^2$ of the beables is given by
\begin{equation}
|\psi({\bf x}_1,\dots,{\bf x}_N,t)|^2  = |\psi_1({\bf x}_1,\dots,{\bf x}_N,t)|^2 + |\psi_2({\bf x}_1,\dots,{\bf x}_N,t)|^2\,.
\label{1.2}
\end{equation}
One can then easily show that for particle beables whose configuration $({\bf x}_1,\dots,{\bf x}_N)$ lie within the support of $\psi_1$ at $t_0$, the guidance equations can be written as 
\begin{equation}
\fr{d {\bf x}_k}{dt}= \frac{1}{m_k} {\boldsymbol {\nabla}}_k S_1\,,
\label{1.3}
\end{equation}
for $t \in I$, where $\psi_1=|\psi_1| \exp(iS_1/\hbar)$. In other words the beables are guided only by the wavefunction $\psi_1$. From time $t_0$ onwards, one can just ignore the wavefunction $\psi_2$ in the description of these particle beables whose configuration $({\bf x}_1,\dots,{\bf x}_N)$ lies within the support of $\psi_1$ at $t_0$. This is what we call an effective collapse. The probability that we have an effective collapse $\psi \to \psi_1$ is given by the probability that the beable configuration $({\bf x}_1,\dots,{\bf x}_N)$ lies within the support of $\psi_1$ at $t_0$. In quantum equilibrium, this probability is given by
\begin{equation}
\int d^3 x_1 \dots d^3 x_N |\psi_1({\bf x}_1,\dots,{\bf x}_N,t)|^2 = \frac{|\langle \psi_1 |\psi \rangle|^2}{||\psi_1||^2}\,.
\label{1.4}
\end{equation}
Hence the effective collapse $\psi \to \psi_1$ occurs with the same probability as the ordinary collapse $\psi \to \psi_1$. Similarly one can have effective collapse to $\psi_2$. One can of course also have effective collapse if the state is a superposition of more than two states.

For all practical purposes, the condition (\ref{1.101}) for having non-overlapping states is in fact too strong. In order to have an effective collapse, it is sufficient to assume that the overlap of $\psi_1$ and $\psi_2$ is minimal and that the tails should be well-behaved, in the sense that the derivatives in the tails should be small enough.

For a superposition of macroscopically distinct states it is clear that we have effective collapse. This is simply because macroscopically distinct systems are located in distinct regions of physical space and hence the corresponding quantum states will be non-overlapping in the configuration space. Decoherence will ensure that, at least for all practical purposes, they remain non-overlapping for all future times. In a quantum measurement-like situation we generally have collapse too. In this case, however, one often needs to include the measurement device in order to see that effective collapse indeed occurs. This was explained in detail already by Bohm in his seminal paper \cite{bohm2}, but it is instructive to review this. 

In a typical measurement situation the system and measurement apparatus are initially described by a product state $(\sum_i c_i\psi^s_i(x_s))\psi^a (x_a)$. The superscript $s$ and $a$ refer hereby respectively to the system and the apparatus. During the measurement process the state evolves into  $\sum_i c_i\psi^s_i(x_s)\psi^a_i(x_a)$ according to the Schr\"odinger evolution. The different wavefunctions of the system $\psi^s_i(x_s)$ in the configuration space $(x_s)$ might be overlapping. But if the wavefunctions $\psi^a_i(x_a)$ are non-overlapping in the configuration space $(x_a)$, i.e.\ $\psi^a_i(x_a)\psi^a_j(x_a)=0, i\neq j$, and remain non-overlapping for all future times, then one has effective collapse, say 
\begin{equation}
\sum_i c_i\psi^s_i(x_s)\psi^a_i(x_a)\rightarrow\psi^s_k(x_s)\psi^a_k(x_a)\,.
\label{2}
\end{equation}
Because the different states $\psi^a_i$ correspond to macroscopically distinct states, for example measurement devices with a macroscopic needle pointing in different directions, it is guaranteed that they will be non-overlapping. 

When an effective collapse occurs in a quantum measurement-like situation, it is guaranteed that the outcome of the measurement is recorded in the beables. That is, the beable configurations contain information about the outcome of the measurement. For example, one can consider the positions of the beables of the macroscopic needle of the measurement device. The beables will then indicate a certain direction which corresponds to the outcome of the measurement. For example, one can have a needle that points up or down depending on whether the spin of the particle was up or down.  If the different wavefunctions of the system $\psi^s_i(x_s)$ have considerable overlap, then the outcome of the measurement is not recorded in the beables of the system, but only in the beables of the measurement device. 

Conversely, one can also say that, if the outcome of a measurement is recorded in the beable configuration in a pilot-wave type model, then effective collapse must have occurred. 

More generally we can see that the requirement that a good pilot-wave model should exhibit effective collapse, at least in situations where one expects ordinary collapse to occur in standard quantum theory, is equivalent to Bell's requirement that the beables should `on the macroscopic level, yield an image of the everyday classical world' \cite[p.\ 41]{bell87}. 

Note that the notion `whenever you expect ordinary collapse to occur in standard quantum theory' is rather vague. Some people would say that collapse occurs when the system is in a superposition of macroscopically distinct states. Other people would end the von Neumann chain later. For example, one could extend the chain to the very end and say that the collapse occurs when the observer makes an observation. Therefore, the requirement that a good pilot-wave model should exhibit effective collapse whenever we have a superposition of macroscopically distinct states is sufficient but not necessary. One could equally well have pilot-wave models in which effective collapse occurs only when observers make observations (see also Bell's footnote 4 \cite[p.\ 41]{bell87}).{\footnote{A problem with Holland's model  \cite{holland,holland881} for fermionic fields is that it never exhibits effective collapse. However, because the model was presented only for free fermions, one could solve this problem by including bosonic fields in the description and by introducing suitable beables for these bosonic degrees of freedom \cite{struyve052}.}}

An additional point to note is that, if one wants the pilot-wave model to exhibit psycho-physical parallelism, then outcomes of measurements also need to be recorded in the beables corresponding to the observer's mental state. Or, in other words, the different wavefunctions corresponding to different mental states, which are correlated to the different outcomes of the measurement, should be non-overlapping (which will then yield effective collapse).

\section{Similar models}\label{similarmodels}
In this section we describe models that are similar to our model in the sense that they do not associate beables with every degree of freedom of the wavefunction. We start with the general idea behind these models.

\subsection{General framework}\label{generalframework}
Suppose we have two Hilbert spaces ${\mathcal H}_i$, $i=1,2$ with bases $B({\mathcal H}_i) = \left\{ |o_{i} \rangle \big| o_{i} \in O_i \right\}$, where the $O_i$ are some index sets. Consider now the product Hilbert space ${\mathcal H}={\mathcal H}_1\otimes{\mathcal H}_2$. The set 
\begin{equation}
B({\mathcal H}_1\otimes{\mathcal H}_2  ) = \left\{ |o_{1}, o_{2} \rangle \bigg| |o_{1}, o_{2} \rangle = |o_{1}\rangle \otimes | o_{2} \rangle;\,  |o_{i}\rangle \in B({\mathcal H}_i),\, i=1,2 \right\}
\label{3}
\end{equation} 
then forms a basis for the product space. In this basis a quantum state $| \psi \rangle$ can be expressed as 
\begin{equation}
| \psi \rangle = \sum_{o_{1}, o_{2}} \psi(o_{1}, o_{2} ) |o_{1}, o_{2} \rangle \,.
\label{3.1}
\end{equation}
The corresponding density matrix reads
\begin{equation}
{\widehat \rho} = | \psi \rangle \langle \psi | = \sum_{^{o_{1}, o_{2}}_{{\bar o}_{1}, {\bar o}_{2 }}} \psi^*({\bar o}_{1}, {\bar o}_{2} ) \psi(o_{1}, o_{2} )  |o_{1}, o_{2} \rangle  \langle {\bar o}_{1}, {\bar o}_{2} |  \,.
\label{3.2}
\end{equation}
In the basis $B({\mathcal H}_1\otimes{\mathcal H}_2)$ the coefficients of the density matrix are 
\begin{equation}
\rho(o_{1}, o_{2};{\bar o}_{1}, {\bar o}_{2})= \psi^*({\bar o}_{1}, {\bar o}_{2} ) \psi(o_{1}, o_{2} ) \,.
\label{3.23}
\end{equation}

Suppose now we want to introduce beables corresponding only to the degree of freedom $o_{1}$. One can do this by considering the reduced density matrix 
\begin{equation}
{\widehat \rho}_1 = \textrm{Tr}_2 {\widehat \rho} = \sum_{{o_{1},{\bar o}_{1}, o_{2}}}  \psi^*({\bar o}_{1}, o_{2} )\psi(o_{1}, o_{2} )  |o_{1} \rangle  \langle {\bar o}_{1} |\,.
\label{3.3}
\end{equation}
In the basis $B({\mathcal H}_1)$ this matrix has coefficients 
\begin{equation}
\rho_1(o_{1};{\bar o}_{1}) = \sum_{o_{2}}  \psi^*({\bar o}_{1}, o_{2} ) \psi(o_{1}, o_{2} ) \,.
\label{3.4}
\end{equation}
The probability of finding the system 1 in the state $|o_{1}\rangle$ is given by $\rho(o_{1})=\rho_1(o_{1};o_{1})$. Potentially one can interpret $\rho(o_{1})$ as a density of beables corresponding to the degree of freedom $o_{1}$. Given the Schr\"odinger equation for $|\psi \rangle$, the velocity field for these beables can then be found by considering the continuity equation for the density $\rho(o_{1})$.

In summary, we obtain a pilot-wave model in two steps. In the first step, we consider the density matrix and we trace out over some degrees of freedom. In the second, we try to find a pilot-wave model starting from this reduced density matrix. Of course, if we want the model to reproduce the quantum predictions, we need to introduce enough beables, so that, as explained in the previous section, the beables yield an image of the everyday classical world. 

We now continue with some models in which this framework has been succesfully applied.

\subsection{Bell's model for non-relativistic spin-1/2 particles}
A first example is Bell's model for non-relativistic spin-1/2 particles. Quantum mechanically, a non-relativistic spin-1/2 particle is described by a state which is an element of the Hilbert space ${\mathcal H}={\mathcal H}_1\otimes{\mathcal H}_2$, where ${\mathcal H}_1$ is the Hilbert space corresponding to square integrable functions on ${\mathbb{R}}^3$ and ${\mathcal H}_2$ is the two-dimensional Hilbert space corresponding to the Pauli $\sigma$-matrix representation of the rotation group $SU(2)$. For ${\mathcal H}_1$ we use the position basis $B({\mathcal H}_1)=\left\{ |{\bf x} \rangle \big| {\bf x} \in {\mathbb{R}}^3  \right\}$ and for ${\mathcal H}_2$ we use the basis 
\begin{equation}
B({\mathcal H}_2) = \left\{ | a \rangle \big| a=-1,1  ;  {\widehat {\sigma}}_3  | a \rangle  =a  | a \rangle    \right\}\,.
\label{3.40001}
\end{equation}
In the product basis 
\begin{equation}
B({\mathcal H})= \left\{|{\bf x}, a \rangle  \big| |{\bf x}, a \rangle = |{\bf x} \rangle \otimes | a \rangle;   |{\bf x} \rangle \in B({\mathcal H}_1);  |a \rangle \in B({\mathcal H}_2) \right\}
\label{3.4001}
\end{equation} 
a state $| \psi (t)\rangle \in {\mathcal H}$ can be expanded as
\begin{equation}
| \psi (t)\rangle = \sum_{a}  \int d^3x \psi_a({\bf x},t)  |{\bf x}, a \rangle \,.
\label{3.5}
\end{equation}
The dynamics for the expansion coefficient $\psi_a({\bf x},t)$ is given by the Pauli equation
\begin{equation}
i\pa_t\psi_a=-\frac{\hbar}{2m}\left(\boldsymbol{\nabla}-\frac{ie}{\hbar c}{\bf A}\right)^2\psi_a + \sum_{b} \mu{{\widehat{\boldsymbol \sigma}}}_{ab}\cdot{\bf B} \psi_b+V\psi_a\,,
\label{3.6}
\end{equation}
with ${\bf A}$ the electromagnetic vector potential, ${\bf B}=\boldsymbol{\nabla}\times{\bf A}$ the corresponding magnetic field and $V$ an additional scalar potential. $\mu$ is the magnetic moment. 

Bell \cite{bell66,bell71,bell82} proposed a pilot-wave model for the non-relativistic spin-1/2 particle by introducing beables only for the position degree of freedom of the wavefunction $\psi_a({\bf x},t)$ and not for the spin degree of freedom. Let us consider how this works in the context of our general framework. 

In the product basis $B({\mathcal H})$, the density matrix $\rho$, corresponding to the state $| \psi \rangle$, has coefficients 
\begin{equation}
\rho_{a;a'}({\bf x};{\bf x}',t) = \psi^*_{a'}({\bf x}',t) \psi_{a}({\bf x},t)\,.
\label{3.7}
\end{equation}
By tracing out over the spin degree of freedom, we obtain the reduced density matrix
\begin{equation}
\rho({\bf x};{\bf x}',t) = \sum_{a}\psi^*_a ({\bf x}',t) \psi_{a}({\bf x},t)\,.
\label{3.8}
\end{equation}
Following Bell, we can now take $\rho({\bf x},t) = \rho({\bf x};{\bf x},t)$ as the density of particle beables. The dynamics for the particle beables is found by considering the continuity equation
\begin{equation}
\pa_t\rho+\boldsymbol{\nabla} \cdot {\bf j}=0\,,
\label{4}
\end{equation}
with
\begin{equation}
{\bf j}=\sum_{a}\left(\fr{\hbar}{2mi}\left(\psi^{*}_a\boldsymbol{\nabla}\psi_a - \psi_a \boldsymbol{\nabla}\psi^{*}_a\right)-\fr{e}{mc}{\bf A}\psi^{*}_a\psi_a\right)\,.
\label{5}
\end{equation}
 The guidance equation is then given by
\begin{equation}
\fr{d {\bf x}}{dt}=\fr{{\bf j}}{\rho}\,.
\label{6}
\end{equation}

In this way we arrive at Bell's model. Although no beables were introduced for the spin degrees of freedom of the wavefunction, this model is empirical equivalent to quantum theory. Results of measurements are generally recorded in `positions of things' so that we have effective collapse whenever we expect ordinary collapse to occur \cite{bell87}. 

Bell's view of beables as structure-less point particles is advocated, by a\-mongst others, D\"urr {\em et al.}\ \cite{durr952} and by Bohm and Hiley \cite[pp.\ 204-230]{bohm5}. This does not exclude other models in which beables are introduced also for the spin degrees of freedom, see e.g.\ Bohm {\em et al.}\ \cite{bohm551,bohm552} and Holland \cite{holland881,holland882,holland}.

\subsection{Bell's model for quantum field theory}
Another example is Bell's model for quantum field theory on a lattice \cite{bell86}. In this model the beables correspond to the fermion numbers on the lattice points. Although not mentioned explicitly, the model can be seen as tracing out over the bosonic degrees of freedom of the quantum state.{\footnote{We thank Sheldon Goldstein and Nino Zangh\`i for pointing this out to us.}} 

Bell further realized that there is nothing unique about the choice of beables. We could have others instead or in addition. Following up on this note of Bell, Goldstein {\em et al}.\ \cite{goldstein04} address, in all generality, the issue of whether one should introduce beables for all particle species, i.e.\ for photons, electrons, quarks, etc. Goldstein {\em et al}.\ conclude that, although experiments do not necessarily discriminate between different approaches, there is as yet no compelling mathematical or physical reason not to introduce beables for all particle species. 

\subsection{The model of Squires and Mackman for relativistic wave equations}
Also Squires and Mackman \cite{squires94} suggest to introduce beables (particle beables) only for fermions and not for bosons, but in the context of relativistic quantum theory. Their motivation to do so is that one can construct a pilot-wave model in terms of particle beables for the spin-1/2 Dirac theory, at least on the first quantized level, but not for bosons, in particular not for photons. The principal reason for this is that there is no natural candidate for a future-causal conserved vector for bosons, which can be interpreted as a particle current. This issue is discussed in detail in \cite{kyprianidis85,holland,bohm5,horton02,struyve051}.

If we put Squires model in the context of our general framework we see that the Hilbert space is the product of the bosonic and the fermionic Hilbert space. The pilot-wave model is then found by integrating out the bosonic degrees of freedom. Squires and Mackman did not present this model explicitly but gave an illustration of these ideas in the context of non-relativistic quantum theory. They considered a two-particle wavefunction and devised a pilot-wave model for one particle by integrating out the degree of freedom of the other particle. This simplified model was earlier discussed by Holland \cite[pp.\ 319-321]{holland} and later reconsidered by Goldstein {\em et al}.\ \cite{goldstein04}.{\footnote{The formulation of a pilot-wave model for a system described by a density matrix in position space, which is not necessarily obtained by tracing out some degree of freedom, is also discussed in \cite{bell80,maroney04,maroney05,durr05}.}} 

\section{Pilot-wave model for QED}\label{themodel}
In this section we present our model for quantum electrodynamics (QED). We first review Bohm's pilot-wave model for the free electromagnetic field. It will then be only a small step from Bohm's model for the free electromagnetic field to our model for QED. 

\subsection{Bohm's pilot-wave model for the free electromagnetic field}
In order to arrive at his pilot-wave model Bohm started from the electromagnetic field quantized in the Coulomb gauge, for which the Hamiltonian is given by{\footnote{In this section we use units in which $\hbar = c =1$.}}
\begin{equation}
{\widehat H}_B = \fr{1}{2} \int d^3 x \left( {\widehat {\boldsymbol \Pi}}^T \cdot {\widehat {\boldsymbol \Pi}}^T - {\widehat {\boldsymbol A}}^T \cdot \nabla^2  {\widehat {\boldsymbol A}}^T    \right)\,.
\label{6.001}
\end{equation}
The fields ${\widehat {\bf A}}^T$ and  ${\widehat {\boldsymbol \Pi}}^T$ are respectively the transversal electromagnetic field and the transversal momentum field, i.e.\ we have ${\boldsymbol \na} \cdot {\widehat {\bf A}}^T ={\boldsymbol \na} \cdot {\widehat {\boldsymbol \Pi}}^T = 0 $. The magnetic field operator reads ${\widehat {\bf B}} = {\boldsymbol \nabla } \times {\widehat {\bf A}}^T$ and the transversal part of the electric field operator reads ${\widehat {\bf E}}^T = - {\widehat {\boldsymbol \Pi}}^T $. The longitudinal part of the electric field operator is zero by the Gauss law $ {\boldsymbol \nabla } \cdot {\widehat {\boldsymbol \Pi}} =0$.

The commutation relations for these operators read 
\begin{equation}
[{\widehat A}^T_i({\bf x}),{\widehat \Pi}^T_{j}({\bf y})] = i \left(\delta_{ij} -  \frac{\partial_i \partial_j }{\nabla^2}\right)\delta({\bf x} - {\bf y}) \,.
\label{6.002}
\end{equation}
The other fundamental commutation relations are zero. A representation for these operators is easily found by using the following Fourier expansion of the field operators
\begin{eqnarray}
{\widehat {\bf A}}^T({\bf x}) &=& \frac{1}{(2\pi)^{3/2}} \sum^2_{l=1} \int d^3 k e^{i {\bf k} \cdot {\bf x}} {\boldsymbol \varepsilon}^l({\bf k}) {\widehat q}_l({\bf k})\,, \nonumber\\  
{\widehat {\boldsymbol \Pi}}^T({\bf x})  &=&  \frac{1}{(2\pi)^{3/2}} \sum^2_{l=1} \int d^3 k e^{-i {\bf k} \cdot {\bf x}} {\boldsymbol \varepsilon}^l({\bf k}) {\widehat \pi}_l({\bf k})\,.
\label{6.003}
\end{eqnarray}
Here $ {\widehat q}_l$ and ${\widehat \pi}_l$ are complex operators in momentum space which satisfy the commutation relations
\begin{equation}
[{\widehat q}_l({\bf k}) ,{\widehat \pi}_{l'}({\bf k}') ] = i \delta_{ll'}\delta({\bf k} - {\bf k}') \,, \quad [{\widehat q}_l({\bf k}) ,{\widehat q}_{l'}({\bf k}') ] = [{\widehat \pi}_l({\bf k}),{\widehat \pi}_{l'}({\bf k}')]= 0\,.
\label{6.00301}
\end{equation}
The vectors ${\boldsymbol \varepsilon}^l({\bf k})$, $l=1,2$ are two real, orthogonal polarization vectors, which we choose to obey the following relations
\begin{eqnarray}
&{\bf k} \cdot {\boldsymbol \varepsilon}^l({\bf k}) = 0\,, & \\
&\sum^2_{l=1} {\varepsilon}^l_i({\bf k}) {\varepsilon}^l_j({\bf k})= \delta_{ij} - \frac{k_i k_j}{k^2}\,,& \\
&{\boldsymbol \varepsilon}^l({\bf k}) = {\boldsymbol \varepsilon}^l(-{\bf k})\,. &
\label{6.004}
\end{eqnarray}
From the last relation and the fact that ${\widehat {\bf A}}^T$ and ${\widehat {\boldsymbol \Pi}}^T$ are Hermitian, we have that ${\widehat q}_l({\bf k}) = {\widehat q}^{\dagger}_l(-{\bf k})$ and ${\widehat \pi}_l({\bf k})={\widehat \pi}^{\dagger}_l(-{\bf k})$.

For Bohm's pilot-wave model we need the functional Schr\"odinger representation. This representation is obtained by choosing the complete set of eigenstates of the operators ${\widehat q}_l$ as the basis of the Hilbert space ${\mathcal H}_B$. This set is given by states that are labelled by pairs of smooth functions $(q_1({\bf k}),q_2({\bf k}))$:{\footnote{In fact the functional Schr\"odinger picture introduced as such is ill-defined. Yet, work of Symanzik and L\"uscher seems to imply that the functional Schr\"odinger picture, as it was introduced here, can be made mathematically well-defined, even for interacting theories, by introducing an extra renormalization constant \cite{symanzik81,luscher85}.}} 
\begin{equation}
B({\mathcal H}_B) =  \left\{ | q_1,q_2 \rangle \bigg| {\widehat q}_{l}({\bf k})  | q_1,q_2 \rangle  =  q_{l}({\bf k})  | q_1,q_2 \rangle,\, l=1 ,2  \right\} \,.
\label{6.005}
\end{equation}

In the basis $B({\mathcal H}_B)$, the operators ${\widehat q}_l({\bf k})$ and ${\widehat \pi}_l({\bf k})$ have the matrix components
\begin{eqnarray}
\langle q_1,q_2  | {\widehat q}_l({\bf k}) | q'_1,q'_2 \rangle  &=& q_l({\bf k})  \delta (q_1 - q'_1)\delta (q_2 - q'_2)    \,, \nonumber\\  
\langle q_1,q_2 | {\widehat \pi}_l({\bf k})| q'_1,q'_2 \rangle  &=& - i \fr{\de }{\de q_l({\bf k})}  \delta (q_1 - q'_1)\delta (q_2 - q'_2)     \,. 
\label{6.005001}
\end{eqnarray}
The components of the Hamiltonian are given by
\begin{equation}
\langle q_1,q_2  | {\widehat H}_B | q'_1,q'_2 \rangle = {\widehat H}_B (q,-i\delta / \delta q) \delta (q_1 - q'_1)\delta (q_2 - q'_2) \,,
\label{6.005002}
\end{equation}
with
\begin{equation}
 {\widehat H}_B (q, -i\delta / \delta q) =  \frac{1}{2} \int d^3 k \left( - \frac{\de^2}{\de q^*_{l}({\bf k}) \de q_{l}({\bf k}) } + k^2   q^*_{l}({\bf k}) q_{l}({\bf k})\right)\,.
\label{6.005003}
\end{equation}

Quantum states have the following expansion
\begin{equation}
| \Psi(t) \rangle = \int {\mathcal D}q_1{\mathcal D}q_2 \Psi(q_1,q_2,t) | q_1,q_2 \rangle \,,
\label{6.006}
\end{equation}
with expansion coefficients $\Psi(q_1,q_2,t)$ which are functionals, called wavefunctionals, defined on the configuration space of fields $(q_1,q_2)$. These are the probability amplitudes to find a quantum system in a certain field configuration. 

The dynamics for these wavefunctionals is given by the following Schr\"odinger equation 
\begin{eqnarray}
i \partial_t \Psi(q_1,q_2,t) &=&  {\widehat H}_B\Psi(q_1,q_2,t) \nonumber\\
&=& \frac{1}{2} \int d^3 k \left(- \frac{\de^2}{\de q^*_{l}({\bf k}) \de q_{l}({\bf k}) }   + k^2 q^*_{l}({\bf k}) q_{l}({\bf k})  \right)\Psi(q_1,q_2,t)\,.
\label{6.007}
\end{eqnarray}

The continuity equation for the field probability density 
\begin{equation}
|\Psi(q_1,q_2,t)|^2 =  |\langle q_1,q_2 | \Psi(t) \rangle |^2
\label{6.00701}
\end{equation}
is given by 
\begin{equation}
\frac{\pa |\Psi|^2}{\pa t} +  \sum^2_{l=1} \int d^3 k  \frac{\delta J_l }{\delta q_l}   =0\,,
\label{6.008}
\end{equation}
with 
\begin{equation}
J_l({\bf k};q_1,q_2,t ) = |\Psi(q_1,q_2,t )|^2 \frac{\delta S(q_1,q_2,t )}{\delta  q^*_l({\bf k})  }  
\label{6.009}
\end{equation}
the field current and $\Psi = |\Psi| \exp(iS)$. The pilot-wave interpretation is obtained by defining the guidance equation
\begin{equation}
{\dot q}_l({\bf k})   = \frac{J_l({\bf k};q_1,q_2,t)}{|\Psi(q_1,q_2,t )|^2}=  \frac{\delta S(q_1,q_2,t )}{\delta  q^*_l({\bf k})  } 
\label{6.00901}
\end{equation}
for the field beables $(q_1,q_2)$. Over an ensemble, the field beables are further assumed to be distributed according to the equilibrium density $|\Psi(q_1,q_2,t)|^2$. This is the pilot-wave interpretation for the electromagnetic field that was originally presented by Bohm \cite{bohm2} and which was further developed by Ka\-lo\-ye\-rou \cite{kaloyerou85,kaloyerou94,kaloyerou96}.{\footnote{Further reviews on pilot-wave theory in terms of field beables can be found in \cite{valentini92,valentini96,valentini04,bohm872,bohm5,holland,holland93,struyve051,struyve052}.}} 

One could argue that the fields $(q_1,q_2)$ are not suitable beables because they live in momentum space. However, the fields $(q_1,q_2)$ are in a one-to-one relations with a transversal vector field ${\bf A}^T$ and a field ${\bf B}$, which live in physical space. These fields are defined by 
\begin{eqnarray}
{\bf A}^T({\bf x}) &=& \frac{1}{(2\pi)^{3/2}} \sum^2_{l=1} \int d^3 k e^{i {\bf k} \cdot {\bf x}} {\boldsymbol \varepsilon}^l({\bf k}) q_l({\bf k})\,, \label{6.010} \\
{\bf B}({\bf x}) &=& {\boldsymbol \nabla} \times {\bf A}^T({\bf x}) =  \frac{i}{(2\pi)^{3/2}} \sum^2_{l=1} \int d^3 k e^{i {\bf k} \cdot {\bf x}} {\bf k}\times {\boldsymbol \varepsilon}^l ({\bf k}) q_l({\bf k})\,.
\label{6.011}
\end{eqnarray}
Hence using the fields $(q_1,q_2)$, which live in momentum space, as beables is equivalent to using the fields ${\bf A}^T$ or ${\bf B}$, which live in physical space, as beables. The field beable ${\bf B}$ is also the field that is revealed in a quantum measurement of the magnetic field. 

For a solution $(q_1(t),q_2(t))$ to the guidance equation, we can also associate a field ${\bf E}^T$, which is given by
\begin{equation}
{\bf E}^T({\bf x},t) = -\partial_t {\bf A}^T({\bf x},t) =  -\frac{1}{(2\pi)^{3/2}} \sum^2_{l=1} \int d^3 k e^{i {\bf k} \cdot {\bf x}} {\boldsymbol \varepsilon}^l({\bf k}) \partial_t q_l({\bf k},t)\,.
\label{21.3}
\end{equation}
Unlike to the field beable ${\bf B}$, we need the time evolution of the beables $(q_1,q_2)$ to construct the field ${\bf E}^T$. In this way, the fields ${\bf B}$ and ${\bf E}^T$ take on roles analogous to position and momentum in non-relativistic quantum theory. Alternatively one could also have developed a pilot-wave model in which the beables correspond to the transversal part of the electric field. 

\subsection{Standard formulation of QED}\label{standard formulation}
Before presenting our pilot-wave model in Subsection \ref{pilot-wavemodel}, we will here recall some features of the standard formulation of QED. In Subsection \ref{quantumpredictions} we will then explain how our pilot-wave model reproduces the quantum predictions.

We start with the formulation of QED in the Coulomb gauge, which can be found in e.g.\ \cite[pp.\ 346-350]{weinberg95}. In the Coulomb gauge, the Hamiltonian reads\begin{equation}
{\widehat H} = {\widehat H}_B + {\widehat H}_F + {\widehat H}_I + {\widehat V}_C\,,
\label{7}
\end{equation}
with ${\widehat H}_B$ the free Hamiltonian for the electromagnetic field which was defined in equation ({\ref{6.001}}), ${\widehat H}_F$ the free Hamiltonian for the Dirac field
\begin{equation}
{\widehat H}_F = \int d^3 x {\widehat \psi}^{\dagger}  \left( -i {\boldsymbol \al} \cdot  {\boldsymbol \na}  \right){\widehat \psi}\,,     
\label{8}
\end{equation}
${\widehat H}_I$ the interaction Hamiltonian
\begin{equation}
{\widehat H}_I = - \int d^3 x  {\widehat {\boldsymbol A}}^T \cdot {\widehat {\bf j}}\,, 
\label{9}
\end{equation}
and ${\widehat V}_C$ the Coulomb potential
\begin{equation}
{\widehat V}_C = \frac{1}{2}\int d^3 x d^3 y \frac{{\widehat j}^0({\bf x}){\widehat j}^0({\bf y})}{4\pi |{\bf x} -{\bf y}|}\,.
\label{10}
\end{equation}
The operator ${\widehat j}^\mu$ is the Dirac charge current
\begin{equation}
{\widehat j}^\mu = e {\widehat \psi}^{\dagger} \gamma^0 \gamma^\mu  {\widehat \psi} =  e \left( {\widehat \psi}^{\dagger} {\widehat \psi} ,   {\widehat \psi}^{\dagger}  {\boldsymbol \al} {\widehat \psi}     \right)\,.
\label{11}
\end{equation}
The commutation relations for the electromagnetic field operators are the same as in the free case. The commutation relations for the fermionic field operators read
\begin{equation}
{\{{\widehat \psi}_a({\bf x}), {\widehat \psi}^{\dagger}_b({\bf y})\} = \delta_{ab} \delta({\bf x} - {\bf y})     }\,.
\label{12}
\end{equation}
The other fundamental commutation relations of the fields are zero.

Because the commutation relations of the electromagnetic field operators are the same as in the free case we can use the same representation for these operators as in the free case. For the fermionic field operators we will not choose an explicit representation because we will integrate out the fermionic degrees of freedom. The Hilbert space for QED is then the direct product Hilbert space of a bosonic Hilbert space ${\mathcal H}_B$ and a fermionic Hilbert space ${\mathcal H}_F$, with ${\mathcal H}_B$ the Hilbert space with basis $B({\mathcal H}_B)$ and ${\mathcal H}_F$ the Hilbert space with basis
\begin{equation}
B({\mathcal H}_F) =  \left\{ | f \rangle \right\}
\label{16}
\end{equation}
which is left unspecified. The label $f$ can be discrete or continuous. 

In the product basis $| q_1,q_2 \rangle \otimes | f \rangle=| q_1,q_2, f \rangle$ of ${\mathcal H}_B \otimes {\mathcal H}_F$, the operators ${\widehat q}_l({\bf k})$ and ${\widehat \pi}_l({\bf k})$ now have matrix components
\begin{eqnarray}
\langle q_1,q_2 , f | {\widehat q}_l({\bf k}) | q'_1,q'_2, f' \rangle  &=& q_l({\bf k})  \delta (q_1 - q'_1)\delta (q_2 - q'_2)  \delta_{ff'}  \,, \nonumber\\  
\langle q_1,q_2 , f | {\widehat \pi}_l({\bf k})| q'_1,q'_2, f' \rangle  &=& - i \fr{\de }{\de q_l({\bf k})}  \delta (q_1 - q'_1)\delta (q_2 - q'_2)  \delta_{ff'}   \,. 
\label{16.1}
\end{eqnarray}
The components of the Hamiltonian will be written as
\begin{equation}
\langle q_1,q_2 , f | {\widehat H} | q'_1,q'_2, f' \rangle = {\widehat H}_{ff'} (q,-i\delta / \delta q) \delta (q_1 - q'_1)\delta (q_2 - q'_2) \,.
\label{16.2}
\end{equation}
For example, the bosonic part of the Hamiltonian now reads
\begin{equation}
\left( {\widehat H}_B \right)_{ff'}(q, -i\delta / \delta q) = \delta_{ff'} \frac{1}{2} \int d^3 k \left( - \frac{\de^2}{\de q^*_{l}({\bf k}) \de q_{l}({\bf k}) } + k^2  q^*_{l}({\bf k}) q_{l}({\bf k})  \right)\,.
\label{16.3}
\end{equation}

A state $| \Psi (t)\rangle \in {\mathcal H}_B \otimes {\mathcal H}_F$ has the expansion coefficients $\Psi_f(q_1,q_2,t)$ in the product basis. So the expansion coefficients are wavefunctionals on the configuration space of fields, just as in the case of the free electromagnetic field, but now they carry an extra label $f$. This label $f$ represents the fermionic degrees of freedom. Note the analogy with spin, where the wavefunction lives on ordinary configuration space and carries a spin-index. Using the notation introduced in (\ref{16.2}) we find that $\Psi_f(q_1,q_2,t)$ satisfies the Schr\"odinger equation
\begin{equation}
i \pa_t\Psi_f(q_1,q_2,t) = \sum_{f'} {\widehat H}_{ff'} (q,-i\delta / \delta q)\Psi_{f'}(q_1,q_2,t)\,.
\label{16.4}
\end{equation}

\subsection{Pilot-wave model for QED}\label{pilot-wavemodel}
We can now construct a pilot-wave model with beables only for the bosonic degrees of freedom, by following the general framework given in Subsection \ref{generalframework}. The density matrix for the state $| \Psi (t)\rangle$ has coefficients
\begin{equation}
\rho_{f;f'} (q_1,q_2;q'_1,q'_2,t) = \Psi^*_{f'}(q'_1,q'_2,t )\Psi_f(q_1,q_2,t )\,.
\label{17}
\end{equation}
By tracing out over the fermionic degrees of freedom we obtain the reduced density matrix
\begin{equation}
\rho(q_1,q_2;q'_1,q'_2,t) = \sum_f  \rho_{f;f} (q_1,q_2;q'_1,q'_2,t) =  \sum_f \Psi^*_f(q'_1,q'_2,t )\Psi_{f}(q_1,q_2,t )\,.
\label{18}
\end{equation}

We take $\rho(q_1,q_2,t) = \rho(q_1,q_2;q_1,q_2,t)$ as the density of the beables $q_1$ and $q_2$. By using the Schr\"odinger equation (\ref{16.4}) we find  
\begin{eqnarray}
\pa_t \rho(q_1,q_2,t) &=& \sum_f \left( \pa_t \Psi^*_f(q_1,q_2,t ) \Psi_{f}(q_1,q_2,t )  + \Psi^*_f(q_1,q_2,t ) \pa_t  \Psi_{f}(q_1,q_2,t ) \right)  \nonumber\\
&=& \sum_{f,f'}  i\Bigg( \left(  {\widehat H}_{ff'}(q, -i\delta / \delta q)   \Psi_{f'}(q_1,q_2,t ) \right)^* \Psi_{f}(q_1,q_2,t )\nonumber\\
&&- \Psi^*_f(q_1,q_2,t ) {\widehat H}_{ff'}(q, -i\delta / \delta q)\Psi_{f'}(q_1,q_2,t ) \Bigg) \,.
\label{18.1}
\end{eqnarray}
In this expression only the kinetic part of the free Hamiltonian for the electromagnetic field survives. This kinetic term has components
\begin{equation}
\left( {\widehat H}^{{\textrm{kin}}}_B \right)_{ff'}(q, -i\delta / \delta q) = -\delta_{ff'}\frac{1}{2} \int d^3 k \frac{\de^2}{\de q^*_{l}({\bf k}) \de q_{l}({\bf k}) }\,.
\label{18.2}
\end{equation}
In this way we find the continuity equation
\begin{eqnarray}
\pa_t \rho(q_1,q_2,t) &=& \frac{1}{2i}  \sum_f \sum^2_{l =1} \int d^3k \Bigg(  \Psi_{f}(q_1,q_2,t ) \frac{\de^2}{\de q^*_{l}({\bf k}) \de q_{l}({\bf k})} \Psi^*_f(q_1,q_2,t )\nonumber\\
&& - \Psi^*_f(q_1,q_2,t ) \frac{\de^2}{\de q^*_{l}({\bf k}) \de q_{l}({\bf k})} \Psi_{f}(q_1,q_2,t )   \Bigg)\nonumber\\
&=& - \sum^2_{l =1} \int d^3k \frac{\de}{\de q_{l}({\bf k})}  J_{{l}}({\bf k};q_1,q_2,t)\,,
\label{19}
\end{eqnarray}
with 
\begin{equation}
J_{{l}}({\bf k};q_1,q_2,t) =\sum_f |\Psi_f(q_1,q_2,t )|^2 \frac{\delta S_f(q_1,q_2,t )}{\delta  q^*_l({\bf k})  } \,, \qquad  l=1,2\,,
\label{20}
\end{equation}
where we have used $\Psi_f= |\Psi_f|\exp(iS_f)$. From the continuity equation we can identify the following guidance equation for the field beables
\begin{equation}
\fr{\pa q_{l}({\bf k},t)}{\pa t} = \frac{J_{{l}}({\bf k};q_1,q_2,t)}{\rho(q_1,q_2,t)} \,, \qquad  l=1,2\,.
\label{21}
\end{equation}

Note that, just in Bohm's model for the free electromagnetic field, we introduced beables only for the transversal degrees of freedom of the vector potential. In particular, we do not introduce beables for the longitudinal degrees of freedom, nor for the scalar degrees of freedom of the electromagnetic field. This implies that we do not have beables corresponding with the charge density of the fermionic field. If we did have beables for the longitudinal degrees of freedom or scalar degrees of freedom of the electromagnetic field, these beables could be related to the charge density through the constraints $j^0 = - \nabla^2 A^0$ and $j^0 = {\boldsymbol \nabla} \cdot {\bf E}={\boldsymbol \nabla} \cdot {\bf E}^L$ \cite[pp.\ 346-350]{weinberg95}.

\subsection{How the pilot-wave model reproduces the quantum predictions}\label{quantumpredictions}
In our pilot-wave model the wavefunctional evolves according to the Schr\"o\-ding\-er equation at all times. There is no collapse and therefore no need to refer to ill-defined notions such as measurements, observers, etc. However, we still need to show that our pilot-wave model reproduces the quantum predictions. A basic requirement is of course that we need to assume the equilibrium distribution for the field beables. But as explained earlier on, we also need to show that we have an {\em effective collapse} when we expect ordinary collapse or stated equivalently, that the beables contain an image of the everyday classical world.

\subsubsection{Effective collapse}
The notion of effective collapse in our model is slightly different compared to the one in non-relativistic quantum theory. Let us first explain this. Suppose we have a system described by a superposition
\begin{equation}
\Psi_f(q_1,q_2,t)=\Psi^{(1)}_f(q_1,q_2,t)+\Psi^{(2)}_f(q_1,q_2,t)\,.
\label{22}
\end{equation}
The quantum states $\Psi^{(1)}_f(q_1,q_2,t)$ and $\Psi^{(2)}_f(q_1,q_2,t)$ are said to be non-overlapping wavefunctionals at time $t_0$ if
\begin{equation}
\Psi^{(1)}_f(q_1,q_2,t_0)\Psi^{(2)}_{f'}(q_1,q_2,t_0) =0\, \quad \forall (q_1,q_2)\,, \, \forall f,f'\,.
\label{23}
\end{equation}
It follows that the density of beables $\rho(q_1,q_2,t_0)$ is given by
\begin{equation}
\rho=\sum_f\left|\Psi_f\right|^2=\sum_f\left|\Psi^{(1)}_f\right|^2+\sum_f\left|\Psi^{(2)}_f\right|^2=\rho^{(1)}+\rho^{(2)}\,,
\label{24}
\end{equation}
at the time $t_0$, with $\rho^{(i)}=\sum_f\left|\Psi^{(i)}_f\right|^2$, $i=1,2$. In addition, we have that 
\begin{equation}
\rho^{(1)}(q_1,q_2,t_0)\rho^{(2)}(q_1,q_2,t_0) =  0\,, \quad \forall (q_1,q_2)\,.
\label{24.1}
\end{equation}
This means that if for example $\rho_1\neq0$ in a region in the configuration space of fields $(q_1,q_2)$, then $\rho_2=0$ in that region, and vice versa. Note that the condition ({\ref{24.1}}) equivalent with the condition ({\ref{23}}).

From (\ref{23}) it further follows that we have a similar decomposition for the current at time $t_0$
\begin{equation}
J_{{l}} = J^{(1)}_{{l}} +J^{(2)}_{{l}}\,, 
\label{24.2}
\end{equation}
with 
\begin{equation}
J^{(i)}_{{l}} = \frac{1}{2i}\sum_f \left( \Psi^{(i)*}_f \frac{\de}{\de q^*_{l}} \Psi^{(i)}_f - \Psi^{(i)}_f \frac{\de}{\de q^*_{l}} \Psi^{(i)*}_f\right)\,,\quad l=1,2
\label{26}
\end{equation}
and $J^{(1)}_{{l}}({\bf k};q_1,q_2,t)J^{(2)}_{{l'}}({\bf k}';q_1,q_2,t) = 0$, $l,l'=1,2$, $\forall {\bf k},{\bf k}'$ and $\forall (q_1,q_2)$. This means that if $J^{(1)}_{{l}'}({\bf k}')\neq0$, for some $l'=1,2$, and some ${\bf k}'$, in a region in the configuration space of fields $(q_1,q_2)$, then $J^{(2)}_{{l}}({\bf k})=0$, $l=1,2$, $\forall {\bf k}$, in that region, and vice versa.

It is now readily seen that, if $\psi_1$ and $\psi_2$ are non-overlapping for the time interval $I=[t_0,+\infty)$, then, for $t \in I$, the velocity  field 
\begin{equation}
\fr{\pa q_l}{\pa t} = \frac{ J_{{l}} }{\rho} = \frac{J^{(1)}_{{l}}  + J^{(2)}_{{l}}    }{\rho^{(1)} +\rho^{(2)}} 
\label{25}
\end{equation}
is given by either $\pa q_l/ \pa t = J^{(1)}_{{l}}/\rho^{(1)} $ or $\pa q_l/ \pa t = J^{(2)}_{{l}}/\rho^{(2)} $.

Hence, if $\psi_1$ and $\psi_2$ are non-overlapping for the time interval $I$, the field beables $q_l({\bf k})$ are always effectively guided by either $\Psi^{(1)}$ or $\Psi^{(2)}$. This means that one can ignore either $\Psi^{(1)}$ or $\Psi^{(2)}$ in the future description of the evolution of the beables. This is what we call an effective collapse. In quantum equilibrium, this probability for effective collapse $\Psi \to \Psi^{(1)}$ is given by $|\langle \Psi^{(1)} | \Psi \rangle|^2/||\Psi^{(1)}||^2 $ and similarly for effective collapse $\Psi \to \Psi^{(2)}$. Hence the probabilities are the same as in standard quantum theory.

As in the case of pilot-wave theory for non-relativistic quantum systems, we can make do with less restrictive conditions for effective collapse. In order to have an effective collapse, it is sufficient to assume that the overlap of $\Psi^{(1)}_f(q_1,q_2,t)$ and $\Psi^{(2)}_f(q_1,q_2,t)$ is minimal and that the tails should be well-behaved, in the sense that the functional derivatives in the tails should be small enough.

\subsubsection{Non-overlapping states}
Essential for effective collapse is that the wavefunctional evolves to a superposition of non-overlapping wavefunctionals. For example, states that correspond to macroscopically distinct classical magnetic fields will be non-overlapping.{\footnote{A clear example of this is provided by coherent states. Coherent states are important in the description of the classical limit of the quantized electromagnetic field and one can explicitly show that coherent states that correspond to different average photon number, to linearly independent momenta, or to different frequency are generally non-overlapping \cite{struyve052}.}} In particular, it is sufficient that the states correspond to magnetic fields that differ only in a certain region of physical space. States that are macroscopically distinct because they correspond to different classical electric fields might correspond to approximately the same classical magnetic field at a certain time and hence may be very much overlapping at that time. Nevertheless, because the states correspond to different classical electric fields, it will be guaranteed that the states become non-overlapping in the near future. This follows from the Ehrenfest relation
\begin{equation}
\frac{\partial \langle {\widehat {\bf B}} \rangle}{\partial t} = -  {\boldsymbol \nabla} \times \langle {\widehat {\bf E}} \rangle\,.
\label{25.001}
\end{equation}
This situation is similar to that in non-relativistic quantum theory, where wavefunctions corresponding to macroscopic systems with the same position but with sufficiently different momenta will become non-overlapping in the near future.

In measurement-like situations the quantum state will generally evolve into a superposition of non-overlapping wavefunctionals. Consider for example the following {\em quantum mechanical} description of a measurement. Suppose we do a measurement on some quantum system and that the outcome of the measurement gets correlated with the direction of some macroscopic needle. In non-relativistic quantum theory there correspond particle beables to the needle so that the outcome of the experiment will be recorded in the particles positions. On the other hand, in our model for QED, there are no beables corresponding to the fermionic degrees of freedom. However, if we continue our quantum description of the experiment, the direction of the macroscopic needle will get correlated with the radiation that is scattered off (or thermally emitted from, etc.)\ the needle. Because these states of radiation will be macroscopically distinct they will be non-overlapping in the configuration space of fields and hence the outcome of the experiment will be recorded in the field beable of the radiation. 
 
It is clear that we have a similar situation in other measurement-like situations. The results of measurement outcomes will become correlated with macroscopically distinct classical states of the electromagnetic field, so that we have effective collapse and a record of the outcome of the experiment in the field beable.  

In pilot-wave theory for non-relativistic quantum theory we had an image of the everyday classical world in the particle beables, because they recorded the positions of macroscopic objects. On the other hand, in our model for QED, we get an image of the everyday classical world in the electromagnetic field beable. In particular, positions of macroscopic objects can be inferred from the electromagnetic field beable.

\section{Conclusion}
We have presented a pilot-wave model for quantum electrodynamics. Beables were introduced only for bosonic degrees of freedom of the quantum state and not for fermionic ones. In addition to field beables corresponding to the degrees of freedom of the electromagnetic field, one could in principle also introduce beables corresponding to the other bosonic fields appearing in the standard model: the (electro-)weak interaction field, strong interaction field and Higgs field. However, in principle, there is no need to do this because the electromagnetic field beables already contain an image of the everyday classical world. 

By introducing beables only for the electromagnetic field, our model is minimalist. However, one could construct models which include beables corresponding also to the fermionic degrees of freedom. For example one could use the particle beables as introduced by Colin or by D\"urr {\em et al.}

\section{Acknowledgements}
We are grateful Owen Maroney, Sebastiano Sonego, Rafael Sorkin and Antony Valentini for discussions. WS is further grateful to Stijn De Weirdt for initial encouragement.

\end{document}